\documentclass[prl,aps,showpacs,floats,twocolumn]{revtex4}

\usepackage{graphicx}
\usepackage{dcolumn}
\usepackage{bm}
\usepackage{amsmath}

\begin{document}

\title{Elastic Convection in Vibrated Viscoplastic Fluids}

\author{Hayato Shiba$^{1,2}$, Jori Ruppert-Felsot$^1$, Yoshiki Takahashi$^1$, Yoshihiro Murayama$^1$, Qi Ouyang$^3$, Masaki Sano$^1$}
\affiliation{$^1$Department of Physics, $^2$Department of Applied Physics, University of Tokyo, Hongo, Bunkyo-ku, Tokyo $^1$113-0033, $^2$113-8656, Japan}
\affiliation{$^3$Department of Physics and Mesoscopic Physics Laboratory, Peking University, Beijing 100871, China}

\date{\today}

\begin{abstract}
    We observe a new type of behavior in a shear thinning yield stress
    fluid: freestanding convection rolls driven by vertical oscillation.
    The convection occurs without the constraint of container boundaries
    yet the diameter of the rolls is spontaneously selected for a wide
    range of parameters. The transition to the convecting state occurs
    without hysteresis when the amplitude of the plate acceleration
    exceeds a critical value. We find that a non-dimensional stress, the
    stress due to the inertia of the fluid normalized by the yield
    stress, governs the onset of the convective motion.

\end{abstract}
\maketitle

\begin{figure}[t]
\centerline{\resizebox{0.44\textwidth}{!}{\includegraphics{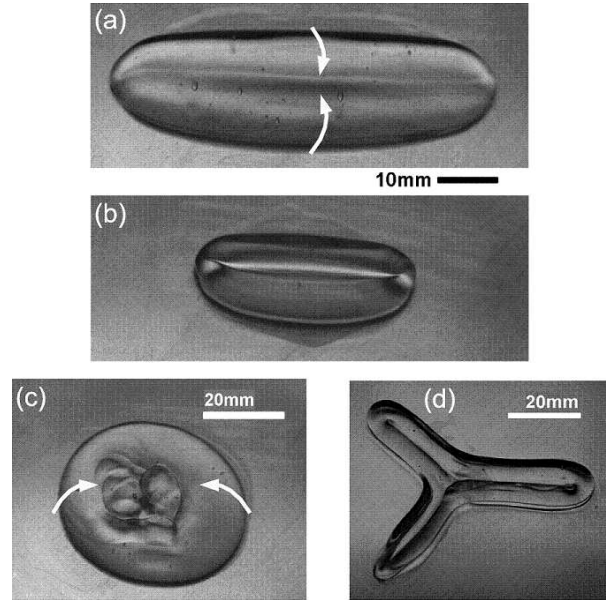}}}
\caption{\label{fig:fig1} (a) Overhead view of the convecting rolls
on a horizontal plate (R1 gel), vertically vibrated at $f=40$ Hz and
acceleration amplitude $\Gamma = 11.1$. The volume of the gel used
was $V=6$ mL. (b) Resulting roll pattern using a smaller volume,
$V=3$ mL, at $f=40$ Hz, $\Gamma =12.0$. The scale bar is the same as
that of (a). (c) Ring pattern obtained at higher volumes. The mean
convecting flow occurs in the outer ring region. $f=40$ Hz, $\Gamma
= 15.0$, and $V=9$ mL. (d) Rolls with a branch can be formed with
a different initial condition. Oscillation parameters
are $f=100$ Hz, $\Gamma = 32.6$, and $V=5$ mL. Arrows represent 
the direction of the flow.}
\end{figure}

Complex fluids comprise a large number of fluids in everyday and industrial usage.
Viscoelastic fluids often exhibit interesting behavior when subjected to mechanical stresses,
such as rod climbing and the open siphon effect\cite{Bingham,Boger}.
Viscoplastic fluids that posses a yield stress can exhibit both
elastic solid-like behavior and fluid-like behavior.
Thus a blob of viscoplastic fluid can retain its shape when laid on
a plate, even if turned upside down, if the stress due to its own
weight does not exceed the yield stress. With the application of a
mechanical shear stress larger than the yield stress, resistance to
the stress abruptly decreases and the fluid starts to flow or falls
in drops. In the present work, we observe a new instability that
arises in soft yield stress fluids driven far from equilibrium that
reveals characteristics particular to plastic flow which is not
observed in various pattern formation in non-Newtonian
fluids\cite{James,Carre} without yield stress.

Vibrated materials often show a variety of patterns such as
Faraday's surface waves in Newtonian fluids, stripe and labyrinthine
patterns and convection of granular
materials\cite{Umbanhowar,Jaeger}, and localized excitation of
shear-thickening fluids\cite{Merkt}. The convection of vibrated
viscoplastic fluids observed in the present work is a new example of
pattern forming instabilities of vibrated materials. In many
pattern-forming systems, the evolution of the pattern can occur on a
timescale much slower than that of the forcing. Understanding how
the slow mode of the pattern dynamics emerges from the fast
excitation mode is a fundamental problem in the nonlinear dynamics
of far from equilibrium systems.

When a viscoplastic fluid placed on a horizontal plate is vertically
vibrated, freestanding convective rolls appear as a lip shape, as
shown in Fig. \ref{fig:fig1(a,b)}. We observe that a pair of rolls
is spontaneously created, and counterrotate with a slow circulation
speed independent of the fast vibrating mode of the plate. The
convection rolls are freely standing on the horizontal surface
without side and top boundary walls. The diameter of the rolls is
spontaneously selected and not determined by the depth of the
container. This is in contrast to thermal convection where the depth
of the fluid/container constrains the wavelength of the convection. The
critical condition for the onset of convection is independent of the
volume and the height of the fluid for a wide range of parameters,
and is determined by a newly defined parameter: a non-dimensional
stress produced by the inertia of the fluid interacting with the
oscillating plate.
\begin{figure}[b]
\centerline{\resizebox{0.45\textwidth}{!}{\includegraphics{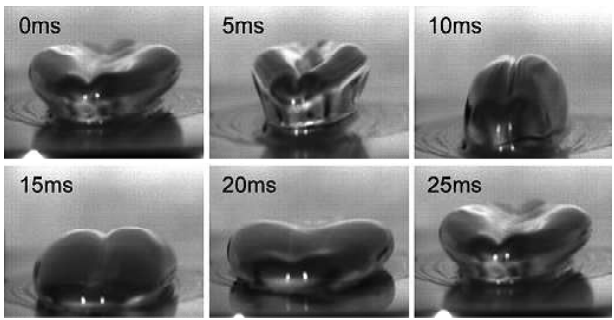}}}
\caption{\label{fig:fig2} Side view of the oscillating motion of the
rolls at various phases of the plate oscillation ($f=40$ Hz, $\Gamma
=14.8$, and $V=5$ mL). Time proceeds from left to right at 5 ms
intervals. The oscillation of the gel is locked to the plate
frequency but out of phase [see Fig. \ref{fig:fig5}(c)]. However,
the mean convective flow occurs on a time scale independent of that
of the oscillation period.}
\end{figure}

The experimental apparatus consists of a horizontal aluminum plate
(120 mm diameter), mounted on a vibration device (VG-100C: Vibration
Test System). The plate is vertically vibrated with a sinusoidal
oscillation with amplitude $A$ and frequency $f$, {\it i.e.}
$z=A\sin (2\pi ft)$, by a function generator (NF 3540). The
dimensionless acceleration amplitude is defined as $\Gamma =A(2\pi
f)^2/g$, where $g$ is the gravitational acceleration. The
acceleration amplitude of the plate is measured by an accelerometer
(PV-90B:RION) to an accuracy of $\pm 0.2$ m/s$^2$. The motion
is recorded by a high speed CMOS camera with a $1024\times
1024$ pixel resolution and time resolution of $1/1000$s. We used
fluids with a monotonically shear-thinning property above the yield
stress strength, {\it i.e.} viscoplastic. We report the results for
a typical shear-thinning fluid gel (sol-gel intermediate), commonly
used for ultrasound examinations (Dane-gel R1, Rohd\'e products).
The material is made of a polymeric gelling agent and water.
However, we observed qualitatively similar results using other shear
thinning yield stress fluids such as xanthan gum, shaving gel, and
toothpaste. We measured the viscosity of the material using a
cone plate rheometer with temperature control (Rheosol G5000, UBM).
We confirmed that the viscosity decreases with increasing shear
rate. We estimated the yield stress of the gel as $\tau_B\simeq 150$
Pa, using the same device by measuring the maximum stress supported
by the gel under constant shear.

A fixed volume of the gel is placed on the plate. The initial shape
of the gel is amorphous. The gel is elastically stretched and
compressed during the plate oscillation period. 
As the plate acceleration is increased
it becomes apparent that the gel is fluidized, and it rapidly
collapses and spreads onto the plate. 
A lump initially laid on the plate is thus
gradually encircled by a rim of inward rotating rolls as the
acceleration is increased. The rim of roll elongates in one
direction along the horizontal surface breaking the azimuthal
symmetry. Within a few seconds the middle part of the lump is
invaded by rolls, and finally become lip-shaped as shown in
Fig.~\ref{fig:fig1} (a-b).

Detailed dynamics can be seen with the use of a high speed camera.
The rolls oscillate elastically at the same frequency as the plate
(shown in Fig.~\ref{fig:fig2}) but with a different phase [shown in
Fig.~\ref{fig:fig5} (a)]. The gel appears to be folding with each
oscillation of the plate. Each roll exhibits inward rotation: outer
edges lift from the surface of the plate and fold toward the center
of the lump to create rolls as shown in Fig. \ref{fig:fig2}. This
motion is very periodic and therefore the net displacement of each
part of the fluid is small after each complete period of the
oscillation.  However, we observe a relatively slow
circulating convection within the rolls, by tracking the motions of
tracers synchronously with the period of the vertical oscillation.
The speed of the convective motion (e.g. $\sim 1$ Hz) is much slower
than the fast mode of the periodic deformation due to the plate
oscillation (e.g. 40 Hz). The motion is independent of the
frequency of oscillation but increases as a function of the
acceleration of the plate above the onset.

\begin{figure}
\centerline{\resizebox{0.45\textwidth}{!}{\includegraphics{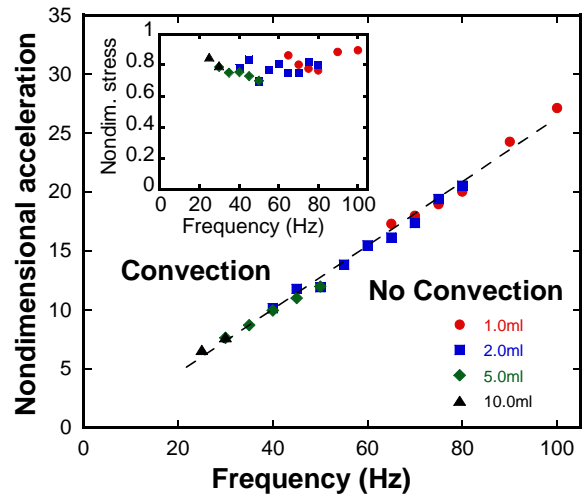}}}
\caption{\label{fig:fig3} State diagram of the materials as a
function of acceleration and frequency. The fitted line shows the
state boundary. Below the line the rolls do not rotate. Above the
line they can rotate if the roll shape has been clearly formed. The
dashed line is obtained by a linear fit. Inset: The same plot for
non-dimensional stress defined by Eq.~(\ref{eqn:ndstress}). }
\end{figure}

We measure the onset acceleration amplitude $\Gamma_c$ to determine
the dynamical state diagram plotted as a function of frequency,
shown in Fig. \ref{fig:fig3}. Once the roll pattern is formed, the
shape of the rolls persists even after the vertical vibration is
stopped due to the high viscosity and yield stress of the gel. If we
decrease $\Gamma$ from a convecting roll state, rotation of the
rolls ceases immediately when $\Gamma$ becomes lower than the
critical value $\Gamma_c$ and starts as we increase $\Gamma$ above
$\Gamma_c$ without hysteresis for the onset. The state boundary
falls on a line {\it i.e.}, $\Gamma\sim (0.27\pm 0.01)f$. Therefore
the convection starts when the velocity amplitude $v=A(2\pi f)$ is
larger than a certain threshold velocity $v_c =0.39$m/s. This
threshold varies with different kinds of testing materials.
Therefore it is reasonable to introduce non-dimensional parameters
in order to clarify the mechanism of the onset of convection. Here
we introduce a non-dimensional parameter, a stress exerted by the
plate due to the inertia of the material normalized by the yield
stress $\sigma_y$, 
\begin{equation}\label{eqn:ndstress}
\Sigma\equiv\frac{\sigma}{\sigma_y} = \frac{\rho A^2(2\pi
f)^2}{\sigma_y}.
\end{equation}
\begin{figure}
\centerline{\resizebox{0.45\textwidth}{!}{\includegraphics{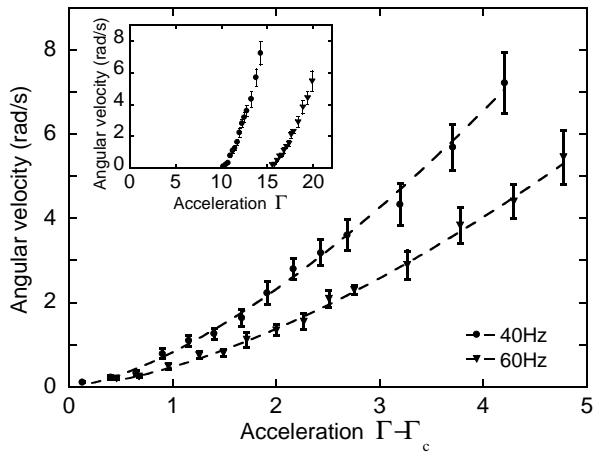}}}
\caption{\label{fig:fig4} The mean angular velocity $\Omega$ as a
function of the acceleration above onset, $\Gamma - \Gamma_c$.
Results for $f=40$ Hz with $V = 5.0$ mL and 60 Hz with $V = 2.2$ mL.
The horizontal axis is shifted such that the
onset value $\Gamma_c$ is at the origin. Each is fit by $\Omega
=\alpha (\Gamma -\Gamma_c)^\beta$, where $\alpha$ and $\beta$ are
fit parameters. This results in $\Omega = (0.82\pm 0.04)\times
(\Gamma -\Gamma_c)^{1.50\pm 0.04}$ [s$^{-1}$] and $\Omega = (0.47\pm
0.02)\times (\Gamma -\Gamma_c)^{1.55\pm 0.03}$ [s$^{-1}$]. Inset:
The same plot without subtracting $\Gamma_c$.}
\end{figure}

The stress due to inertia is estimated to be $\sigma =\rho A\cdot
A(2\pi f)^2 =\rho v^2$, where $\rho$ is the mass density of the
material, $\Sigma$ is the ratio of the exerted stress and the yield
stress. We plot the non-dimensional stress $\Sigma$ for various
volumes of material and various frequencies, in the inset of
Fig.~\ref{fig:fig4}. The critical condition for the onset of
convection is $\Sigma\sim O(1)$. Despite the fact that the
estimation of the yield stress has some ambiguity and the stress
distribution is not uniform, the data for various different
conditions cluster near the line corresponding to $\Sigma =0.8$. The
result strongly supports the hypothesis that the convection starts
when the stress distribution in the material exceeds the yield
stress $\sigma > \sigma_y$, {\it i.e.} above the solid to fluid
transition.

We next measure how the mean angular velocity of the convection
rolls changes above the onset. Rotation of the rolls is monitored by
following small bubbles or density matched tracer particles which
are immersed in the gel for visualization. The convection speed of
the fluid depends on the location within the convecting rolls. The
rolls rotate faster in the middle part and slower at the ends. The
mean angular velocity (radian/sec) near the mid point of the rolls
is obtained from the measurement of the time interval required for
several complete cycles of the tracers. The result is shown in
Fig.~\ref{fig:fig4}. The errorbars in Fig.~\ref{fig:fig4} are
relatively large due to the intrinsic angular velocity distributions
in the rolls as well as statistical variation. We find that the mean
angular velocity increases as an increasing function of the plate
acceleration $\Gamma$ above the onset and is proportional to
$(\Gamma -\Gamma_c)^{1.53\pm 0.03}$.

\begin{figure}[b]
  \includegraphics[width=0.9\linewidth]{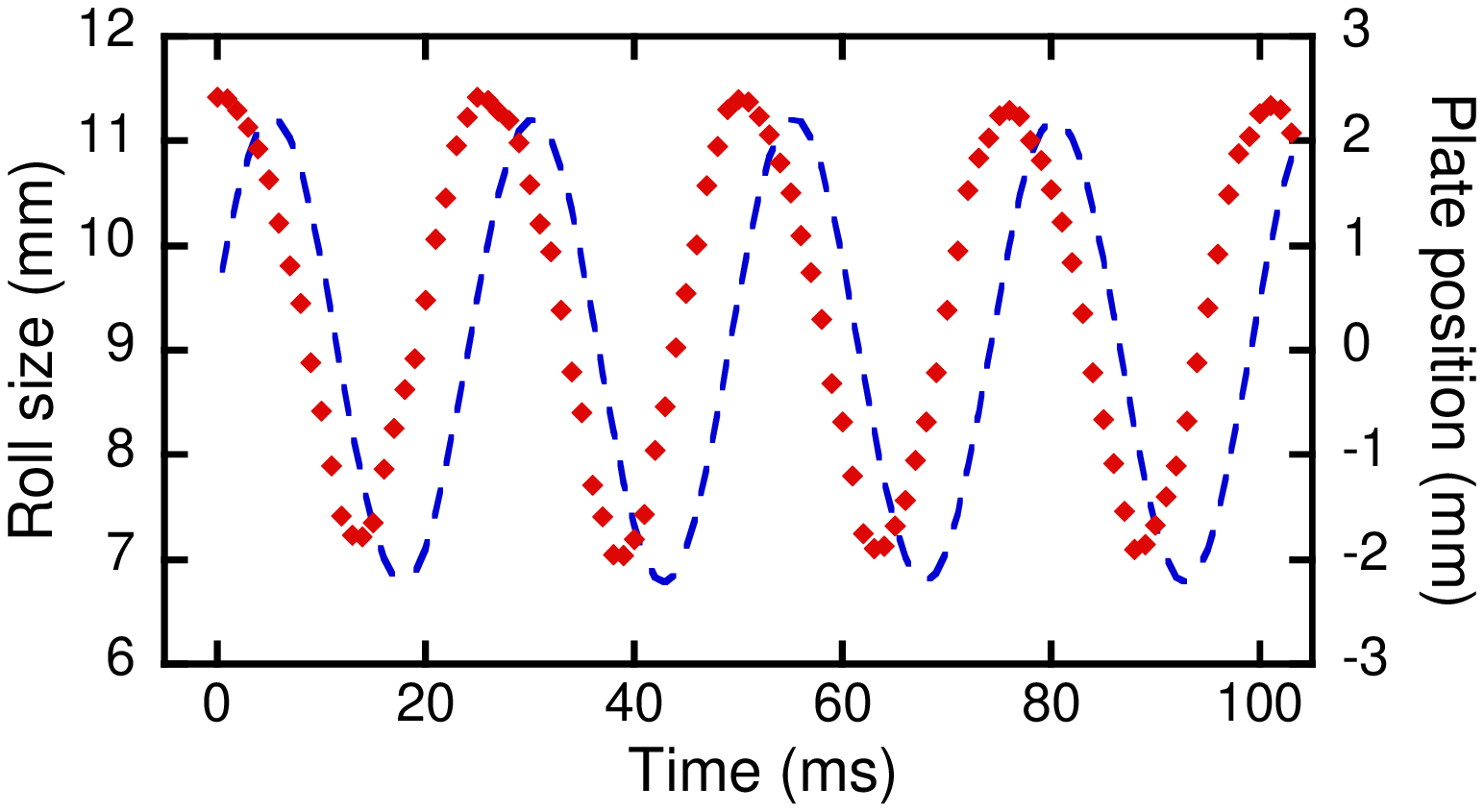}\\
  \includegraphics[width=0.48\linewidth]{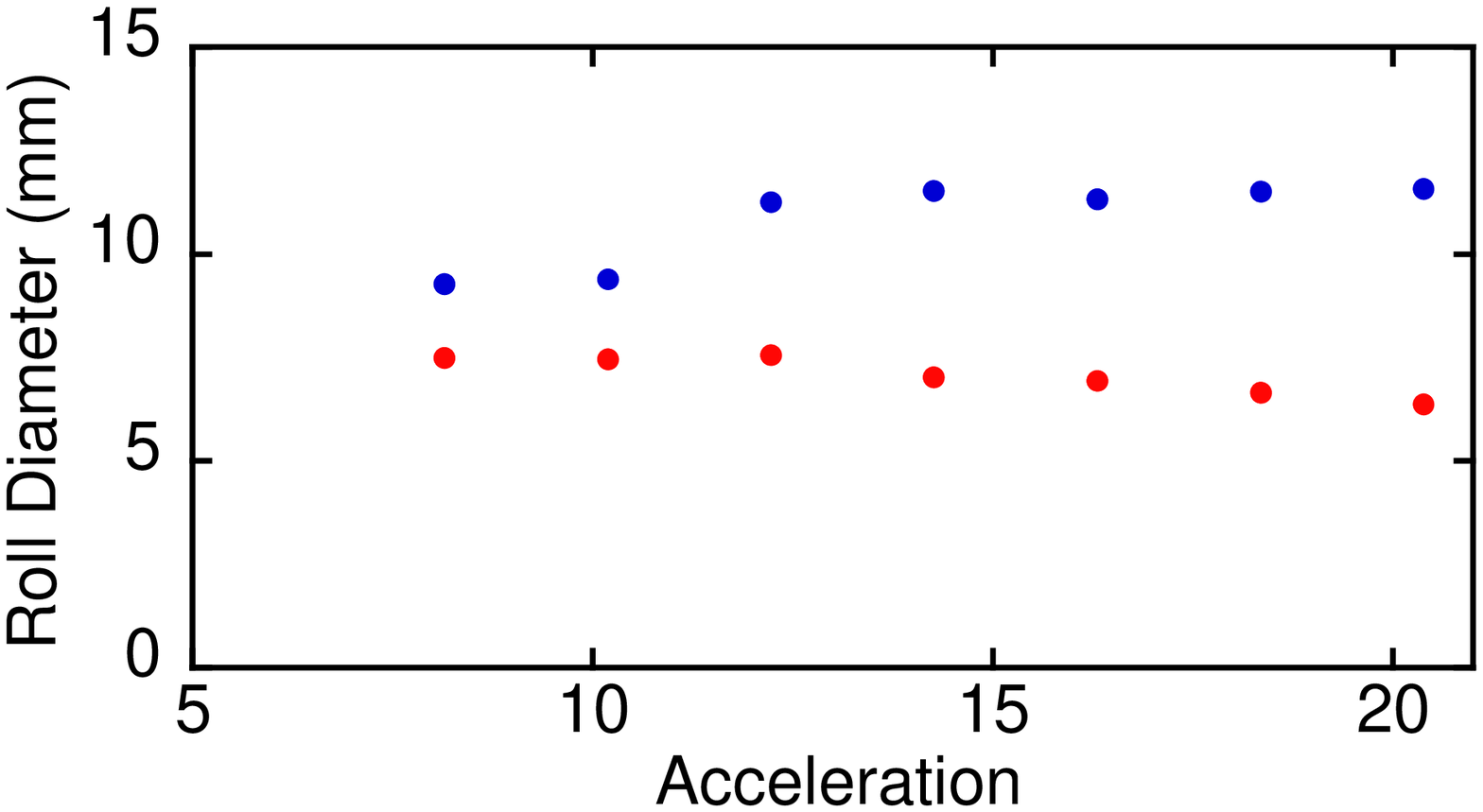}
  \includegraphics[width=0.48\linewidth]{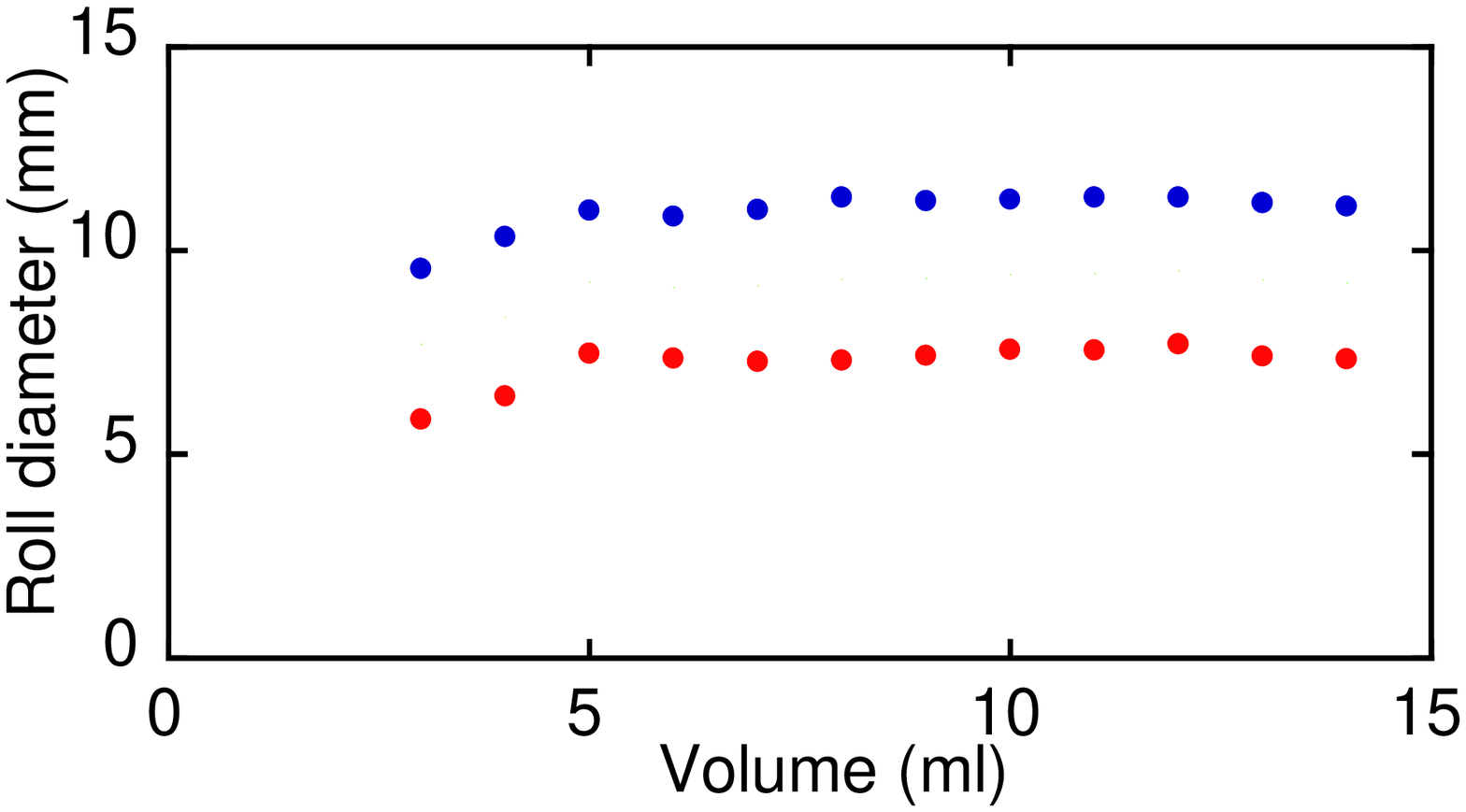}\\
\caption{(a) Dotted line: Time evolution of the diameter of a single
roll as observed from above, looking down onto the plate, for $f=40$
Hz, $V=5$ mL, and $\Gamma =14.3$. Solid line: Position of the plate
showing the phase difference. (b) Dependence of the range of roll
diameters on acceleration $\Gamma$ for $f=40$ Hz, $V=5$ mL
($\Gamma_C=10.1$). The upper (lower) dots show the maximum (minimum)
value of the measured diameter. Note that there is a plateau region
for the maximum of the diameters when the acceleration $\Gamma$ is
well above the onset value $\Gamma_C$. (c) Size dependence of the
roll on the volume of the material measured for $f=40$ Hz,
$\Gamma=12.3$ (above $\Gamma_C$). \label{fig:fig5}}
\end{figure}

Now we note an important feature of the convecting rolls: The
diameter, not the length, of the convective rolls is spontaneously
selected. This is particularly true when the frequency of the plate
is relatively low. {\it i.e.} 30Hz -- 40Hz. Figure~\ref{fig:fig5}(a)
shows the time evolution of the diameter of one roll observed from
above using the high speed camera. We obtain Fig.~\ref{fig:fig5}
(b-c) by plotting the maximum and the minimum observed roll diameter
dependence on acceleration and volume. The maximum and the minimum
values for the diameter of the rolls are independent of acceleration
and volume well above the onset, and for sufficient volume of the
material at lower frequencies, e.g. $f \sim 40$ Hz. For higher
frequencies, e.g. $f \sim 80$ Hz, the diameter of the rolls is
affected by factors such as the initial conditions, suggesting that
the stability band of the wavelength selection is wider for higher
frequency. We can observe very narrow rolls for the same volume of
the material compared with that at low frequencies. We also note
that pattern selection has some initial condition dependence as is
typical of other pattern forming phenomena far from equilibrium. For
example, a triangular vertex of roll pair or a concentric ring of
roll were observed as shown in Fig.~\ref{fig:fig1}(c-d), by imposing
different initial conditions.

\begin{figure}[t]
\centerline{\resizebox{0.45\textwidth}{!}{\includegraphics{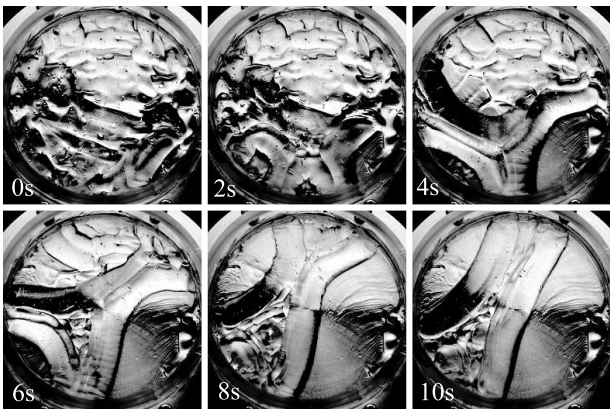}}}
\caption{\label{fig:fig6} Aggregation of the gel in a vertically
vibrated cylindrical container (acrylic resin walls and aluminum
plates). Time proceeds from left to right at intervals of 2 s. The
gel eventually gathers into one area enclosed by rolls and the wall,
vacating the other areas. The experimental parameters are fixed at
$f=50$ Hz, $\Gamma =27.3$, and $V=25$ mL.}
\end{figure}

The measurements described until this point were conducted on a
freestanding blob of gel on a horizontal plate without sidewalls.
However, convection of the gel still occurs in the presence of
sidewalls. When placed in a container, the gel self-segregates into
a ``material sea'' region and a ``dry'' region, with convection
rolls forming the ``shores''. This process is shown in Fig.
\ref{fig:fig6}. An initially uniform shallow (depth of about 15-20
mm) layer of the gel was placed covering the bottom plate of the
cylinder. As the plate acceleration is increased above the critical
acceleration, a roll eventually forms near the wall. Once a roll
forms, it aligns itself perpendicular to the point of contact with
the wall (see Fig. \ref{fig:fig6}). The roll tends to straighten and
thus the action of the roll is to peel the gel from the bottom plate
and move it away from the wall. A very thin layer of gel remains
stuck to the plate in the vacated region. Eventually the gel is thus
segregated into two regions: a shallow region bounded by the
container wall and rolls, and a much thinner layer of gel in the
region vacated by the rolls.

The loss of stability and subsequent transition to the convecting
state of a vertically oscillated viscoplastic fluid occur due to
the solid to fluid transition, as evidenced by the order unity
transition of the nondimensional stress. Above the onset, we suspect that a
nonlinear mechanism such as acoustic streaming may be responsible
for the emergence of the slow mode from the fast oscillating
mode{\cite{Lighthill,Chang}. It is reported that the circulating
motion around a vibrating rod in viscoelastic materials is in the
opposite direction compared to that in Newtonian fluids. The fact
that the flow on the center line is directed toward the rod
coincides with our observation. However, the observed velocity of
our convection and that of acoustic streaming behave very
differently as a function of vibration velocity.
We also speculate that the anisotropy of viscosity plays a crucial role in this phenomenon. Decrease of viscosity (shear thinning) depends on the shear stress tensor. Since the distribution of the stress is anisotropic, the viscosity coefficients of the fluid are also anisotropic. Once the convection rolls are formed by symmetry breaking, the shear in the azimuthal direction is larger than the longitudinal direction. This results in a smaller viscosity in azimuthal direction compared with the longitudinal one, and this effect may stabilize the motions along the azimuthal directions and straighten the rolls.

The newly found state, freestanding convection, can become a new
demonstration for viscoplasticity, the presence of a yield stress in
a complex fluid. Similar phenomena can be expected to occur in
materials ranging from yield stress fluids to amorphous
solids\cite{Lemaitre,Falk,Picard} and glassy
materials\cite{Coussot,Varnik}. Furthermore, enhanced mixing due to
stretching and folding of the fluids during convection may allow for
applications to mixing of such fluids.

We thank T. Kataoka, K. Ito, and M. Shibayama for use of their rheometer, 
K. Takeuchi and S. Tatsumi for useful discussion and experimental help. 
This work was supported by a Japanese Grant-in-Aid for Scientific Research 
from Ministry of Education, Culture, Science, and Technology.



\begin{thebibliography}{99}
\bibitem{Bingham} E.C. Bingham, {\it Fluidity and Plasticity} (McGRAW-HILL Inc., New York, 1922).
\bibitem{Boger} D.V. Boger and K. Walters, {\it Rheological Phenomena in Focus} (Elsevier , Amsterdam, 1993).
\bibitem{James} D.F. James, Nature {\bf 212}, 754 (1966).
\bibitem{Carre} A. Carr\'e, J.-C. Gastel, and M.E.R. Shanahan, Nature {\bf 379}, 432 (1996).
\bibitem{Umbanhowar} P.B. Umbanhowar, F. Melo, and H.L. Swinney, Nature {\bf 382}, 793 (1996).
\bibitem{Jaeger} H.M. Jaeger, S.R. Nagel, and R.P. Behringer, Rev. Mod. Phys. {\bf 68}, 1259 (1996), and references therein.
\bibitem{Merkt} F.S. Merkt, R.D. Deegan, D.I. Goldman, E.C. Rericha, and H.L.Swinney, Phys. Rev. Lett. {\bf 92}, 184501 (2004).
\bibitem{Lighthill} Sir J. Lighthill, J. Sound Vib. {\bf 61}, 391 (1978).
\bibitem{Chang} C. Chang and W.R. Schowalter, Nature {\bf 252}, 686 (1974).
\bibitem{Lemaitre} A. Lemaitre, Phys. Rev. Lett. {\bf 89}, 195503 (2002).
\bibitem{Falk} M.L. Falk and J.S. Langer, Phys. Rev. E {\bf 57}, 7192 (1998).
\bibitem{Picard} G. Picard, A. Ajdari, F. Lequeux, and L. Bocquet, Phys. Rev. E {\bf 71}, 010501R (2005).
\bibitem{Coussot} P. Coussot {\it et al.}, Phys. Rev. Lett. {\bf 88}, 218301 (2002).
\bibitem{Varnik} F. Varnik, L. Bocquet, J.-L. Barrat, and L. Berthier, Phys. Rev. Lett. {\bf 90}, 095702 (2003).
\bibitem{Movie} The movies of the experiments are available at \texttt{http://} \texttt{daisy.phys.s.u-tokyo.ac.jp/EC/}.
\end{thebibliography}
\end{document}